# A Common Semantic Model of the GDPR Register of Processing Activities

Paul RYAN [ac,1] and Harshvardhan J. PANDIT [b] and Rob BRENNAN [a]
[a] *ADAPT, School of Computing, Dublin City University, Dublin 9, Ireland*
[b] *ADAPT, Trinity College Dublin, Dublin 2, Ireland*
[c] *Uniphar PLC, Dublin, Ireland*

**Abstract.** The creation and maintenance of a Register of Processing Activities (ROPA) is an essential process for the demonstration of GDPR compliance. We analyse ROPA templates from six EU Data Protection Regulators and show that template scope and granularity vary widely between jurisdictions. We then propose a flexible, consolidated data model for consistent processing of ROPAs (CSM-ROPA). We analyse the extent that the Data Privacy Vocabulary (DPV) can be used to express CSM-ROPA. We find that it does not directly address modelling ROPAs, and so needs additional concept definitions. We provide a mapping of our CSM-ROPA to an extension of the Data Privacy Vocabulary.

## 1. Introduction

A Register of Processing Activities (ROPA) is a comprehensive record of the personal data processing activities of an organisation. It is central to meet the principle of accountability as set out in Article 30 of the GDPR. Organisations most commonly create and maintain ROPAs through informal tools and spreadsheets[2]. EU Data Protection Regulators also seem to encourage this practice by providing spreadsheet-based templates to assist organisations in preparing and maintaining ROPAs. A spreadsheet, while being a simple and commonly utilised versatile medium, requires effort to enter information and keep it updated. As a human-oriented application, spreadsheets often lack the rich data structures and semantics that are suitable for building automated toolchains, especially when modelling complex legal concepts. The creation of a common data model is required to represent ROPA information across different compliance-related processes and act as the connection between an organisation's internal compliance data and what regulators would expect. This model can be used to fuse information stored in spreadsheets and facilitate the interconnectivity of data processing systems with ROPA-maintenance/compliance systems, automatically update spreadsheets and automated querying, validation, monitoring and reporting of ROPA information. Regulator template consolidation into a semantic model will facilitate an organisation to regulator interoperability; and will provide a single data model for compliance across jurisdictions. The variation of ROPA templates, allied with the option for organisations to develop their own data structures creates significant challenges when it comes to compliance automation and tool development. It is possible to resolve this variation with a flexible, unified data model of a ROPA. It could support multi-jurisdiction tool development for ROPA maintenance and RegTech-style automated compliance reporting to regulators, thus reducing costs [1].

---

[1] Corresponding Author: Paul Ryan, Email: paul.ryan76@mail.dcu.ie . This work is partially supported by Uniphar PLC. and the ADAPT Centre for Digital Content Technology which is funded under the SFI Research Centre's Programme (Grant 13/RC/2106) and is co-funded under the European Regional Development Fund.
[2] IAPP. https://iapp.org/resources/article/measuring-privacy-operations/.

Our research aims to enable the creation of technical solutions for the maintenance of ROPAs through semantic web technologies. We show in this paper that for ROPAs, there are differences within the templates provided by each regulator in terms of semantics and granularity - despite being based on common requirements of GDPR's Article 30. There is existing work regarding the use of semantic vocabularies to represent GDPR for various compliance-related tasks. We select DPV [3] [2] as the most comprehensive and representative vocabulary of the Sot A and answer the research question "to what extent can the existing Data Privacy Vocabulary (DPV) be extended to build a semantic ROPA model spanning the range of regulator ROPA templates". To address these issues, we first consolidate the different regulator issued templates into a Common Semantic Model for ROPAs (CSM-ROPA). We then evaluate and extend the DPV for representing CSM-ROPA. The contributions of this paper are (i) analysis of six ROPA templates from EU data protection regulators (ii) a consolidated semantic model of ROPA and (iii) extensions of the DPV for representing a semantic model of a ROPA. The rest of the paper is structured as follows: Section 2 presents an analysis of the ROPA templates provided by EU regulators. In Section 3, we will present our Semantic Model of a ROPA, and provide an evaluation of the DPV to represent CSM-ROPA.

## 2. Analysis of ROPA templates from EU Data Protection Regulators

We evaluated 6 ROPA templates provided by EU Data Protection Regulators from the jurisdictions of Belgium, Cyprus, Denmark, Finland, Luxembourg and the United Kingdom, selected for their use of the English language. Each was evaluated in terms of file format, the number of fields, relationship with GDPR Article 30 requirements fields, and controlled vocabularies provided. Our analysis also considered guidance documents or pre-populated samples provided by regulators. We found that all six templates meet the minimum GDPR Article 30 requirements by containing the 12 mandatory information fields it requires, but there was variation in the way they modelled each field. The key differences between the templates arise from the extent of data gathered through the information fields. Three ROPA templates (Finland, Denmark and Luxembourg) are direct transcriptions from Article 30 of the GDPR, containing only the 12 prescribed input fields. The other regulators' (Belgium, United Kingdom and Cyprus) ROPA templates have additional information requirements with varying complexity of the information required. The Belgian ROPA also contains a detailed controlled vocabulary of potential inputs for some fields. In the next step in our analysis, we carried out a systematic review of the concepts included in the six templates. We identified synonyms, overlapping and related concepts. We made direct relationships such as composition or qualifications such as domain and range that were implicit in the spreadsheets. We derived 43 unique concepts representing a consolidated ROPA template covering all six jurisdictions. Based on the interpretation of the GDPR and the use of concepts in ROPA, we combined these 43 concepts into the UML model represented in Figure 1.

---

[3] https://w3.org/ns/dpv

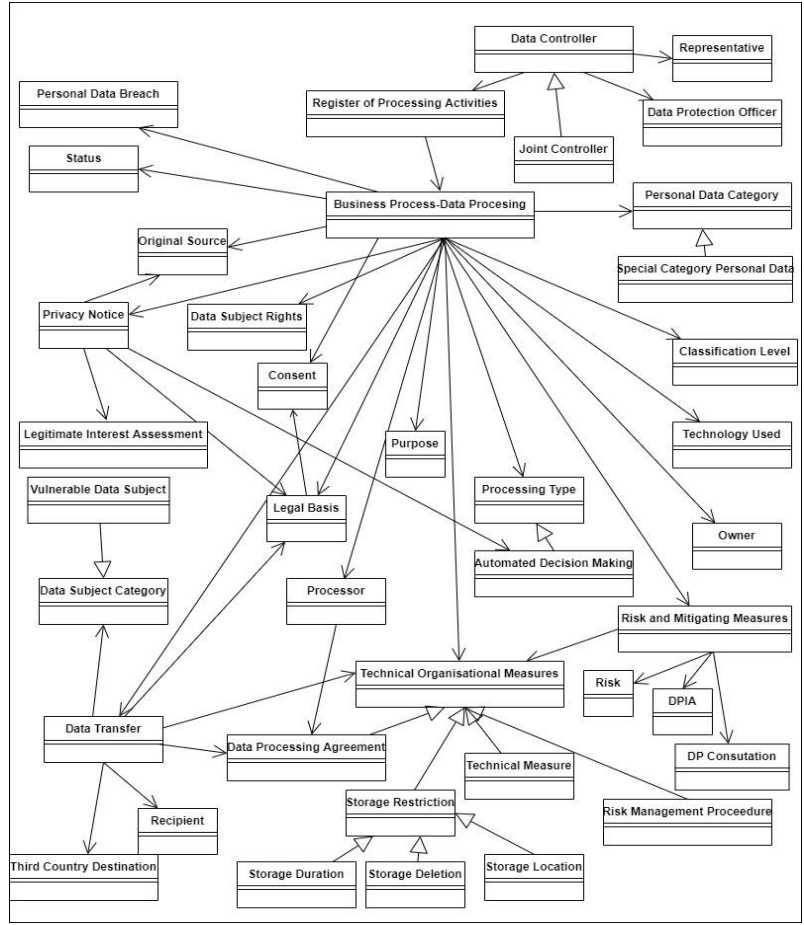

Figure 1. UML Representation of the Combined ROPA Model based on Templates Provided by EU Data Protection Regulators

### 3. A Common Semantic Model for the ROPA

In order to build the semantic ROPA model, we identified the Data Privacy Vocabulary (DPV)[3] as the most relevant and suitable resource to map our ROPA due to its status as a community specification through the W3C Data Privacy Vocabulary and Controls Community Group (DPVCG). The first task was to establish to what extent DPV could represent the combined ROPA model. We compared the suitability of terms in DPV for representing the 43 unique concepts identified from ROPA templates. Table 1 presents an example of the mapping process. Please refer to [4] for full table is available and analysis. We categorised the mapping as "Exact" if the field exactly corresponds to an existing DPV concept indicating no change required. If the data field has a corresponding concept in the DPV that requires an extension to DPV, we categorised it as a 'Partial' match. If the required field can be specified using a combination of multiple concepts in DPV, we categorised the match as 'Complex'. If the concept is missing and needs to be added to the DPV, we categorised the match as 'None' (see Table 1). We found that 14 of our 43 identified unique fields had exact matches with DPV, 15 had partial matches, 3 had complex matches, whereas 11 unique fields had no match within DPV. Thus, the DPV requires the addition of 11 concepts and extension of 15 existing concepts in order to represent information required by the ROPA templates. The additional concepts required are International Transfers, Controller Contact Details, Original Source of Data, Data Protection Officer, Data Protection Impact Assessment, Data Subject Rights, Risk, Privacy Notice, Representative & Data Breach.

**Table 1**. Extract Taken from Mapping Table ROPA Unique Fields to DPV Concepts [4]

| GDPR Regulation | Combined ROPA Model Field | Required GDPR Art. 30 | Related DPV Concept | DPV mapping outcome | Combined No. of Specified Field Values vs DPV properties |
|---|---|---|---|---|---|
| 30.1(b) | Purposes of Processing | Y | dpv: Purpose | Exact | 65 /33 |
| 13/14/15 | Data Subject Rights | N | No DPV Concept | None | |
| 44-47 | Transfer to Third Country | N | dpv:LegalBasis | Partial | |

Most ROPA templates did not suggest any relationships for ROPA fields. Only 7 of the 43 unique fields specified any properties for ROPA fields. These properties were matched against the DPV. The results are displayed in Table 1 in the column titled "Combined No. of Specified Field Values vs DPV". The DPV will require additional expressiveness here in the form of additional properties to meet the requirements of the ROPA[4]. Alternatively, these additional properties, such as "address" can be met through other standardised vocabularies such as vCard.

**Conclusion**

Our research analysed six English language ROPA templates provided by EU Data Protection Regulators in terms of information required and relation with requirements of GDPR's. We identified 43 unique concepts to represent a consolidated common model that enables the representation of ROPAs that span multiple jurisdictions. We then evaluated the DPV as representative of the State of the Art and found that it can currently represent only 32 of the 43 concepts of the common semantic model ROPA (CSM-ROPA). We developed an extension to the DPV with missing concepts and a profile of simple and complex correspondences to DPV. In our future work, we will incorporate our work with the DPV standardisation process. We have provided our work to the Data Privacy Vocabulary Community Group [2] (see footnote 4).

[4] https://lists.w3.org/Archives/Public/public-dpvcg/2020Sep/0006.html